\documentclass{emulateapj}
\shorttitle{Cooling Accretion Disk of XTE~J1817$-$330}
\shortauthors{Rykoff et al.}

\begin{document}

\newcommand\xte{XTE~J1817$-$330}
\newcommand\swift{\emph{Swift}}

\title{\swift{} Observations of the Cooling Accretion Disk of \xte}

\author{
E.~S.~Rykoff\altaffilmark{1},
J.~M.~Miller\altaffilmark{2},
D.~Steeghs\altaffilmark{3},
M.~A.~P.~Torres\altaffilmark{3}
}

\altaffiltext{1}{Department of Physics, University of Michigan, 450 Church St.,
Ann Arbor, MI, 48109, erykoff@umich.edu}
\altaffiltext{2}{Department of Astronomy, University of Michigan, 500 Church
  St., Ann Arbor, MI 48109}
\altaffiltext{3}{Harvard-Smithsonian Center for Astrophysics, 60 Garden St.,
  Cambridge, MA 02138}

\begin{abstract}

The black hole candidate X-ray transient \xte{} was observed by the \swift{}
satellite over 160 days of its 2006 outburst with the XRT and UVOT instruments.
At the start of the observations, the XRT spectra show that the 0.6-10 keV
emission is dominated by an optically thick, geometrically thin accretion disk
with an inner disk temperature of $\sim0.8$ keV, indicating that the source was
in a high/soft state during the initial outburst phase.  We tracked the source
through its decline into the low/hard state with the accretion disk cooling to
$\sim 0.2\,\mathrm{keV}$ and the inner disk radius consistent with the
innermost stable circular orbit at all times.  Furthermore, the X-ray
luminosity roughly follows $L_X \propto T^4$ during the decline, consistent
with a geometrically stable blackbody.  These results are the strongest
evidence yet obtained that accretion disks do not automatically recede after a
state transition, down to accretion rates as low as $0.001 L_{Edd}$.
Meanwhile, the near-UV flux does not track the X-ray disk flux, and is well in
excess of what is predicted if the near-UV emission is from viscous dissipation
in the outer disk.  The strong correlation between the hard X-ray flux and the
near-UV flux, which scale as $L_X^{0.5}$, indicate that reprocessed emission is
most likely the dominate contribution to the near-UV flux. We discuss our
results within the context of accretion disks and the overall accretion flow
geometry in accreting black holes.

\end{abstract}


\section{Introduction}

At high mass accretion rates (those corresponding to $L_{X}/L_{Edd.}  \geq
0.1$, approximately), it is expected that standard geometrically-thin but
optically thick accretion disks should drive accretion onto compact objects.
Stellar-mass black holes with low-mass companion stars provide an excellent
laboratory for the study of disks.  Many of these systems are transient, and
the dramatic flux variations and state changes can be understood as resulting
from variations in the accretion rate.  The timescales in such disks are
readily accessible (the viscous timescale for the entire disk is on the order
of weeks), and light from the accretion process should strongly dominate that
from the companion star.  Fits to X-ray spectra with thermal disk models
reveal trends broadly consistent with $L \propto T^{4}$ at high accretion
rates~\citep*[e.g., ][]{mfm04}.

Although it has long been assumed that light from the accretion disk dominates
the emission from infrared to soft X-rays in actively accreting stellar mass
black holes, recent observations suggest that this may not always be the case.
For instance, \citet{hbmbn05} find that the infrared emission in the
stellar-mass black hole GX~339$-$4 may be jet-based synchrotron radiation.  In
the stellar-mass black hole XTE~J1118$+$480, \citet{hrpgc06} find that even the
UV emission may be partially due to synchrotron emission.  On the other hand,
\citet{rfhbh06} showed that the optical/near-infrared (NIR) emission from an
ensemble of 33 black hole X-ray binaries (BHXRBs) is more consistent with
reprocessed hard X-ray emission rather than synchrotron emission or direct
observations of emission from the accretion disk.

It is also expected that accretion disks should be radially truncated at low
mass accretion rates (corresponding to $L_{X}/L_{Edd.} \leq 0.01$), and
replaced by an advection-dominated flow~\citep*{ecn97}.  Recent analyses of
GX~339$-$4, Cygnus X-1, and SWIFT J1753.5$-$0127 have revealed cool accretion
disks that appear to remain close to the black hole at low accretion
rates~\citep{mhsrh06,mhm06}.  However, observations which strongly constrain
the nature of radio to X-ray emission mechanisms and the nature of accretion disks
in stellar-mass black holes are still relatively few.

Dense multi-wavelength sampling has proved to be very powerful in revealing the
nature of accretion flows in compact objects.  Some of the most productive
studies have correlated X-ray and radio flux measurements in black holes, as a
coarse probe of disk--jet coupling~\citep[see, e.g.,][]{gfp03,mhd05}.  New work
has linked specific (perhaps disk--driven) timing signatures in X-ray power
spectra with radio jet flux~\citep*{mfv05}.  The large number of
contemporaneous X-ray and radio observations made during the outbursts of
transient black holes has been central to this work, and has been greatly
facilitated by the ability of Rossi X-ray Timing Explorer (\emph{RXTE}) to make
frequent observations.

The \swift{} Gamma-Ray Burst Explorer~\citep{gcgmn04} is dedicated to the
discovery and follow-up study of gamma-ray bursts.  It also offers the
extraordinary opportunity to perform new multi-wavelength studies of X-ray
transients that can reveal the nature of accretion disks around compact objects
and the relation between disks, hard X-ray coronae, and radio jets.  \swift{}
carries the X-ray Telescope~\citep[XRT,][]{bhnkw05}, an imaging CCD
spectrometer that reaches down to 0.3~keV.  This provides substantially
improved coverage of thermal accretion disk spectra over {\it RXTE}, which has
an effective lower energy bound of 3~keV.  Moreover, the {\it Swift} optical/UV
telescope \citep[UVOT,][]{rkmna05} is ideally suited to tracing inflow through
the accretion disk (UV emission should lead X-ray emission in a standard thin
accretion disk), and to testing the origin of near-ultraviolet (NUV) emission
in accreting black holes.

\xte{} was discovered as a new bright X-ray transient with the \emph{RXTE}
All-sky monitor (ASM) on 2006 January 26~\citep{rlmms06}.  The ASM hardness
ratios suggested a very soft source, making \xte{} a strong black hole
candidate~\citep[e.g.][]{wm84}.  Initial pointed observations with {\it RXTE}
confirmed the soft spectrum, and hinted at very low absorption along the line
of sight~\citep{mhst06}.  A likely radio counterpart was detected on 2006
January 31~\citep{rdm06}, followed by the detection of counterparts at NIR,
optical and ultraviolet wavelengths~\citep{dgccm06,tsjlm06,stmj06}.  Pointed
X-ray observations with \emph{Chandra} confirmed that the absorption along the
line of sight to \xte{} is very low~\citep{mhsw06}.  The NUV
apparent magnitude of \xte{} was comparable to the optical and infrared
magnitudes, which further supports a low reddening towards the source.

The low absorption along the line of sight to \xte{} and its relatively simple
fast-rise, exponential-decay (FRED) outburst profile made it an excellent
source for intensive study with {\it Swift}.  Through approximately 160 days of
the 2006 outburst, {\it Swift} made a total of 21 snapshot observations in the
NUV and X-ray bands.  The resulting data represent the best sample of
simultaneous NUV and X-ray follow-up observations yet obtained from a
stellar-mass black hole evolving from the high/soft state to the low/hard
state.  Herein, we present an analysis of these simultaneous X-ray and NUV
observations of \xte.  In Section 2, we detail the methods used to reduce and
analyze the data.  In Section 3, we present our analysis and results, including
a re-analysis of \swift{} observations of GRO~J1655$-$40.  Our conclusions
follow a brief summary in Section 4.

\section{Observations and Data Reduction}

\begin{figure}
\begin{center}
\scalebox{0.7}{\rotatebox{270}{\plotone{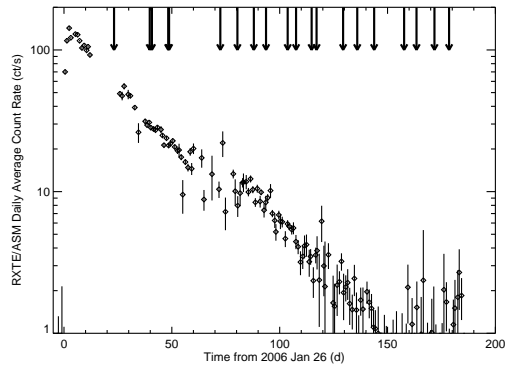}}}
\caption{\label{fig:asmswift}\emph{RXTE}/ASM light curve of the 2006 outburst
  of \xte.  The count rates are daily averages in the sum (1.5-12 keV) band.
  The arrows denote the dates of the \swift{} observations, as described in
  Table~\ref{tab:xrayobs}.
}
\end{center}
\end{figure}

\swift{} visited \xte{} on 21 occasions between 2006 February 18 and 2006 July
23.  Tables~\ref{tab:xrayobs} and \ref{tab:uvotobs} provide a journal of these
observations.  Figure~\ref{fig:asmswift} shows the \emph{RXTE}/ASM light curve in
the sum band (1.5-12 keV), with arrows indicating the dates of the \swift{}
observations.  XRT observations were taken in windowed timing (WT) mode while
the transient was very bright, and in photon counting (PC) mode for the final
four observations.  The UVOT exposures were taken in six filters ($V$, $B$,
$U$, $UVW1$, $UVW2$, $UVM2$) for the first three observations, and with the
$UVW1$ filter subsequently.  Table~\ref{tab:uvotobs} provides the effective
wavelengths for these filters.

\begin{deluxetable*}{ccccc}
\tablewidth{0pt}
\tablecaption{XRT Observation Log\label{tab:xrayobs}}
\tabletypesize{\scriptsize}
\tablehead{
\colhead{Obs.} & \colhead{Start} &
\colhead{Exposure Time} & \colhead{Count Rate\tablenotemark{c}} &
\colhead{Exclusion Box/Radius}\\
Number\tablenotemark{a,b} & (UT) & (s) & ($\mathrm{ct}\,\mathrm{s}^{-1}$) & (pixels)
}
\startdata
001 & 2006 02 18 04:50:34 & 1105 & 776 & 15\\

002 & 2006 03 05 19:12:02 & 1467 & 544 & 10\\
    & 2006 03 06 17:57:47 & 530  & 532 & 10\\

003 & 2006 03 07 16:25:32 & 669 & 516 & 10\\
    & 2006 03 07 18:01:32 & 805 & 514 & 10\\

004 & 2006 03 15 07:26:31 & 1650 & 427 & 10\\

005 & 2006 03 15 16:59:36 & 1000 & 430 & 10\\

006 & 2006 04 08 11:25:05 & 835  & 266 & 5\\

007 & 2006 04 16 07:49:04 & 598 & 292 & 5\\
    & 2006 04 16 09:26:08 & 590 & 286 & 5 \\

008 & 2006 04 24 00:20:08 & 890 & 259 & 5\\

009 & 2006 04 29 12:33:40 & 145 & 210 & 3\\
    & 2006 04 29 14:10:40 & 325 & 199 & 3\\
    & 2006 04 29 15:46:40 & 740 & 178 & 3\\

010 & 2006 05 09 16:41:46 & 975 & 165 & 3\\

011 & 2006 05 13 13:56:13 & 825 & 150 & 2\\

012 & 2006 05 20 19:24:02 & 955 & 120 & 2\\

013 & 2006 05 23 00:30:44 & 619 & 117 & 2\\
    & 2006 05 23 02:09:27 & 690 & 107 & 2\\

015 & 2006 06 04 07:50:45 & 560 & 77 & 0\\
    & 2006 06 04 09:27:45 & 375 & 77 & 0\\

016 & 2006 06 10 21:18:37 & 746 & 49 & 0\\
    & 2006 06 10 22:55:59 & 720 & 45 & 0\\

017 & 2006 06 18 17:24:44 & 684 & 24 & 0\\
    & 2006 06 18 19:03:40 & 560 & 25 & 0\\

018 & 2006 06 27 13:28:02 & 302 & 5 & 0\\

019 & 2006 07 02 15:34:23 & 519 & 1.8 & 3\\
    & 2006 07 02 17:11:23 & 455 & 1.9 & 3\\

020 & 2006 07 08 10:09:08 & 890 & 2.6 & 5\\

021 & 2006 07 16 17:11:50 & 434 & 1.59 & 3\\
    & 2006 07 16 18:46:37 & 440 & 1.55 & 3\\

022 & 2006 07 23 11:42:36 & 628 & 4.6 & 7\\
    & 2006 07 23 13:19:35 & 624 & 3.9 & 7\\

\enddata
\tablenotetext{a}{There was no \swift{} observation 014.}
\tablenotetext{b}{Observations 001-018 were taken in windowed timing (WT)
  mode.  Observations 019-022 were taken in photon counting (PC) mode.}
\tablenotetext{c}{0.6-10 keV count rates corrected for background, but not
  corrected for pile-up.}
\end{deluxetable*}

\begin{deluxetable*}{ccccccc}
\tablewidth{0pt}
\tablecaption{UVOT Observation Log and Photometry\label{tab:uvotobs}}
\tabletypesize{\scriptsize}
\tablehead{
\colhead{Obs.} & \colhead{Start} & \colhead{Stop} &
\colhead{Net Exp. Time\tablenotemark{a}} & \colhead{Filter\tablenotemark{b}} & 
\colhead{Mag.} & \colhead{$f_\nu$}\\
 & (UT) & (UT) & (s) & & & (mJy)
}
\startdata
001 & 2006 02 18 04:50:39 & 04:53:42 & 183 & $UVW2$ & 14.9(3) & 1.0(3)\\
    & 2006 02 18 04:53:47 & 04:56:54 & 82 & $V$ & 15.2(1) & 2.5(2)\\
    & 2006 02 18 04:56:54 & 04:59:57 & 183 & $UVM2$ &  15.0(1) & 0.9(2)\\
    & 2006 02 18 05:00:03 & 05:03:11 & 102 & $UVW1$ & 14.4(1) & 1.7(1)\\
    & 2006 02 18 05:03:11 & 05:06:20 & 11 & $U$ &  14.4(1) & 2.3(2) \\
    & 2006 02 18 05:06:20 & 05:09:00 & 5 & $B$ & 15.7(2) & 1.9(4)\\
002 & 2006 03 05 19:12:07 & 18:04:35\tablenotemark{c} & 881 & $UVW2$ & 15.3(3) & 0.7(2)\\
    & 2006 03 05 19:20:10 & 18:06:21\tablenotemark{c} & 216 & $V$ & 15.7(1) & 1.7(1) \\
    & 2006 03 05 19:22:13 & 19:28:11 & 358 & $UVM2$ & 15.4(1) & 0.6(1)\\
    & 2006 03 05 19:28:17 & 19:32:16 & 235 & $UVW1$ & 14.7(1) & 1.2(1)\\
    & 2006 03 05 19:32:21 & 19:34:20 & 117 & $U$ & 14.7(1) & 1.8(1)\\
    & 2006 03 05 19:34:25 & 19:36:00 & 93 & $B$ & 15.8(1) & 1.8(2)\\
003 & 2006 03 07 16:25:37 & 18:06:04 & 515 & $UVW2$ & 15.4(3) & 0.7(2)\\
    & 2006 03 07 16:29:50 & 18:07:16 & 127 & $V$ & 15.6(1) & 1.8(1)\\
    & 2006 03 07 16:30:57 & 18:10:41 & 386 & $UVM2$ & 15.4(1) & 0.6(1)\\
    & 2006 03 07 16:34:08 & 18:13:01 & 253 & $UVW1$ & 14.7(1) & 1.2(1)\\
    & 2006 03 07 16:36:17 & 18:14:12 & 89 & $U$ & 14.8(1) & 1.6(1)\\
    & 2006 03 07 18:14:17 & 18:15:00 & 42 & $B$ & 15.8(1) & 1.9(2)\\
004 & 2006 03 15 07:26:36 & 07:54:01 & 1612 & $UVW1$ & 14.9(1) & 1.0(1)\\
005 & 2006 03 15 16:59:42 & 17:06:45 & 423 & $UVW2$ & 15.5(3) & 0.6(2)\\
    & 2006 03 15 17:06:50 & 17:13:53 & 416 & $V$ & 15.7(1) & 1.7(1) \\
    & 2006 03 15 17:13:59 & 17:16:16 & 136 & $UVM2$ & 15.7(1) & 0.5(1)\\
007 & 2006 04 16 07:49:08 & 09:36:00 & 1161 & $UVW1$ & 15.2(1) & 0.76(5)\\
008 & 2006 04 24 00:20:12 & 00:35:00 & 874 & $UVW1$ & 15.4(1) & 0.64(4)\\
009 & 2006 04 29 12:33:45 & 15:59:00 & 1167 & $UVW1$ & 15.6(1) & 0.55(4)\\
010 & 2006 05 09 16:41:51 & 16:58:00 & 954 & $UVW1$ & 15.8(1) & 0.43(3)\\
011 & 2006 05 13 13:56:18 & 14:10:00 & 809 & $UVW1$ & 15.8(1) & 0.43(3)\\
012 & 2006 05 20 19:24:07 & 19:40:00 & 938 & $UVW1$ & 16.0(1) & 0.38(3)\\
013 & 2006 05 23 00:30:50 & 02:21:00 & 1277 & $UVW1$ & 16.1(1) & 0.35(2)\\
016 & 2006 06 10 21:18:42 & 23:08:00 & 1390 & $UVW1$ & 16.4(1) & 0.26(2)\\
017 & 2006 06 18 19:03:45 & 19:13:01 & 547 & $UVW1$ & 16.6(1) & 0.22(2)\\
018 & 2006 06 27 13:28:07 & 16:51:00 & 874 & $UVW1$ & 16.9(1) & 0.17(1)\\
019 & 2006 07 02 15:34:28 & 17:19:01 & 950 & $UVW1$ & 17.1(1) & 0.13(1)\\
020 & 2006 07 08 10:09:13 & 10:24:00 & 873 & $UVW1$ & 17.1(1) & 0.14(1)\\
021 & 2006 07 16 17:11:55 & 18:54:00 & 850 & $UVW1$ & 17.1(1) & 0.13(1)\\
022 & 2006 07 23 11:42:42 & 13:30:00 & 1219 & $UVW1$ & 17.1(1) & 0.14(1)\\
\enddata
\tablenotetext{a}{Net exposure time is the total exposure time with the given
  filter between the designated start and stop times.  Each observation may
  contain more than one exposure with the same filter.}
\tablenotetext{b}{The UVOT filters peak sensitivity is at the following
  wavelengths: $V$ (5460~\AA); $B$ (4350~\AA); $U$ (3450~\AA); $UVW1$
  (2600~\AA); $UVM2$ (2200~\AA); and $UVW2$ (1930~\AA).}
\tablenotetext{c}{Time is on the following day.}
\end{deluxetable*}

\subsection{XRT Data Reduction}

The XRT observations were processed using the packages and tools available in
HEASOFT version 6.1\footnote{See http://heasarc.gsfc.nasa.gov/docs/software/lheasoft}.  Initial event cleaning was performed with ``xrtpipeline''
using standard quality cuts, and event grades 0-2 in WT mode (0-12 in PC mode).
For the WT mode data, source extraction was performed with ``xselect'' in a
rectangular box 20 pixels wide and 60 pixels long.  Background extraction was
performed with a box 20 pixels wide and 60 pixel long far from the source
region.  Several \swift{} observations in WT mode contain multiple pointings
separated by a few hours.  For these observations, each individual pointing was
processed separately, as the detector response varies depending on the location
of the source in the field of view.  For the PC mode data, source extraction
was performed with a 30 pixel radius circular aperture, and background
extraction was performed with an annulus with an inner (outer) radius of 50
(100) pixels.  The PC mode observations with multiple pointings were combined,
in spite of the slight variation in the spectral response, to increase the
signal-to-noise due to the faintness of the source.

After event selection, exposure maps were generated with ``xrtexpomap'', and
ancillary response function (arf) files with ``xrtmkarf''.  The latest response
files (v008) were used from the CALDB database.  All spectra considered in this
paper were grouped to require at least 20 counts per bin using the ftool
``grppha'' to ensure valid results using $\chi^2$ statistical analysis.  The
spectra were analyzed using XSPEC version 11.3.2~\citep{a96}.  Fits made with
the WT mode data were restricted to the 0.6-10 keV range due to calibration
uncertainties at energies less than 0.6 keV\footnote{See
http://swift.gsfc.nasa.gov/docs/heasarc/caldb/swift/docs/xrt/SWIFT-XRT-CALDB-09.pdf}.
Similarly, fits to the PC mode data were restricted to the 0.3-10 keV range.
All of the X-ray flux measurements, unless otherwise noted, are in the 0.6-10
keV range.  The uncertainties reported in this work are 90\% confidence errors,
obtained by allowing all fit parameters to vary simultaneously.

The WT observations were strongly affected by pile-up, especially in the early
observations when the raw uncorrected count rate was as large as
$775\,\mathrm{ct}\,\mathrm{s}^{-1}$ (0.6-10~keV).  When the observations
suffer from pile-up, multiple soft photons can be observed at nearly the same
time, and appear as a single hard photon.  This creates an artificial deficit
of soft photons and an excess of hard photons, and thus hardens the resulting
spectral fit.  To correct for pile-up, we followed the spectral fitting method
described in \citet{rcccc06}: we used various exclusion regions at the center
of the source and re-fit the continuum spectrum (see below) with each exclusion
region.  When changing the exclusion region did not vary the fit parameters
significantly ($<1\sigma$) we had properly corrected for pile-up.  The sizes of
the exclusion regions used are listed in Table~\ref{tab:xrayobs}.  They are
consistent with or larger than those used in \citet{rcccc06}.

\subsection{\label{sec:uvreduction}UVOT Data Reduction}

The UVOT images were initially processed at HEASARC using the standard \swift{}
``uvotpipeline'' procedure, with standard event cleaning.  The initial
astrometric solution of UVOT images is typically offset by 5-10''.  We
corrected for this offset by matching the detected stars with the USNO B1.0
catalog, improving the aspect solution to better than 1''.  After this
correction, we stacked the images in observations with more than one exposure
with the ``uvotimsum'' procedure.

The UVOT images were very crowded due to the low Galactic latitude
($-8.0^\circ$) and relatively low absorption in the line of sight.  This
presented a number of challenges in obtaining an absolute flux calibration for
the NUV images, and we could not use the standard UVOT analysis procedure.
Initial photometry was performed using ``uvotdetect'' and ``uvotmag''.  The
first program, ``uvotdetect'', runs SExtractor~\citep{ba96} to extract sources
and to calculate uncorrected count rates in a fixed aperture.  SExtractor
provides robust background estimations which are useful for this crowded field
where it is difficult to find a source-free background annulus.  The second
program, ``uvotmag'', corrects for coincidence loss and converts count rates to
magnitudes using the photometric zero-points determined in orbit.  The
zero-points were determined with a 12 pixel ($6\arcsec$) radius aperture for
the optical filters ($U$, $B$, $V$) and a 24 pixel ($12\arcsec$) radius
aperture for the NUV filters ($UVW1$, $UVW2$, $UVM2$).  The counterpart to
\xte{} was sufficiently isolated from neighboring stars, and we were able to
use a 12 pixel radius aperture to perform aperture photometry in all
filters. The large aperture sacrifices some signal-to-noise in exchange for
proper coincidence loss correction in the initial observations when the
counterpart is very bright.  In addition, this aperture is matched to that used
to obtain the photometric zero-points of the optical filters, thus avoiding the
need for aperture corrections for these filters.

We were required to perform aperture correction on the images taken with the
NUV filters to use the proper photometric zero-point, which was determined
with a 24 pixel aperture.  For this correction we chose 20 of the most isolated
stars within $5\arcmin$ of the counterpart.  The magnitudes of the
comparison stars were calculated within 12/24 pixel apertures to calculate the
median aperture correction.  The RMS error for this correction was typically
$5-10\%$.  We also confirmed that the corrected light curves of the comparison
stars were stable within $\sim1\%$.  The final magnitudes and flux densities
are shown in Table~\ref{tab:uvotobs}.  The quoted flux density errors include
the errors in the photometric zero-points.

\section{Analysis \& Results}
\subsection{\label{sec:nhred}Column Density and Reddening}

The Galactic column density and reddening in the direction of \xte{} is not
well constrained.  \citet{mhsw06} obtained \emph{Chandra} observations of
\xte{} during the bright outburst, and found a low equivalent neutral hydrogen
column density of $N_H = 8.8-9.7\times10^{20}\,\mathrm{cm}^{-2}$.  This is
smaller than that obtained from the Galactic HI map, which is
$1.6\times10^{21}\,\mathrm{cm}^{-2}$~\citep{dl90}, and implies a relatively
short distance to \xte{}.  \citet{tsmgb06} obtained an optical spectrum during
the early phase of the outburst and estimated the Galactic column density and
reddening using the equivalent width (EW) of several interstellar bands.  They
find the Galactic column density to be between
$1\times10^{21}\,\mathrm{cm}^{-2} < N_H < 3\times10^{21}\,\mathrm{cm}^{-2}$,
and the reddening between $0.1 < E(B-V) < 0.5$, depending on the calibration
chosen.  Given the uncertainties in the $E(B-V)/EW$ relationships for
interstellar bands, we have chosen to use the Galactic column density $N_H$ to
constrain the reddening in the direction of \xte{}.  In any case, the low
column density and low reddening make \xte{} an advantageous target for both
soft X-ray spectroscopy and NUV observations.

After fitting a number of XRT spectra with different continuum models, a value
of $N_H = 1.2 \times 10^{21}\,\mathrm{cm}^{-2}$ was found to yield acceptable
fits with the greatest frequency.  Given that this value is broadly consistent
with other values reported elsewhere, and given that a different choice of
$N_H$ would only introduce an offset in the measured disk properties, and not a
change in the trends reported in Section~\ref{sec:spectra}, $N_H = 1.2 \times
10^{21}\,\mathrm{cm}^{-2}$ was adopted throughout this work.  This column
density implies a reddening of $E(B-V) = 0.215$,~\citep{z90} which is also
consistent with values reported elsewhere; similarly, varying $E(B-V)$ by
$\sim50\%$ does not change the trends described in Section~\ref{sec:simul}.

\subsection{\label{sec:spectra}XRT Spectral Analysis}

We used XSPEC to make spectral fits for each individual XRT observation of
\xte.  Observation 018 was not used because there were insufficient counts to
constrain the spectrum.  As previously mentioned, several WT mode observations
contained multiple pointings which were processed separately.  We fit these
multiple pointings simultaneously, although we did not fix the relative
normalizations, as the overall X-ray flux varied up to $\sim10\%$ on timescales
of less than a day.  To constrain the properties of the cool disk, we have
assumed that the column density to \xte{} does not vary with time, and it has
been fixed at $N_H = 1.2\times10^{21}\,\mathrm{cm}^{-2}$ as described in the
previous section.

The XRT spectra of \xte{} cannot be adequately fit with a continuum model with
a single component.  In the early observations, the soft disk emission
dominates, but there is a significant hard excess.  In the late observations,
the hard X-ray emission dominates, but there is a significant soft excess over
a simple power-law spectrum.  To demonstrate that our conclusions about the
disk emission do not depend on our assumptions about the origin of the hard
X-ray emission we have fit the XRT spectra with two separate two-component
absorbed models.  In each case we start with a simple optically thick
geometrically thin accretion disk with multiple blackbody components, as
described in ``diskbb''~\citep{mikmm84}.  This is combined with (a) a simple
power-law component or (b) a hot optically thin Comptonizing
corona~\citep{t94}.  In all cases the two-component absorbed continuum model is
sufficient to describe the spectra.  In the final observations, when one would
expect to see a relativistically broadened iron line around
$\sim5\,\mathrm{keV}$~\citep{mr06} there are insufficient statistics to allow a
significant constraint on this component.

The ``diskbb'' model is very simple in that it does not factor in relativistic
effects or radiative transfer, nor does it specify a zero-torque condition at
the inner boundary~\citep[see, e.g.,][]{znmm05}.  More sophisticated disk
models have certainly been developed, but ``diskbb'' is well understood, and
correction factors aiming to describe spectral hardening through radiative
transfer are based on this model.  Furthermore, more sophisticated models do
not give statistically superior results.  As we are primarily interested in
trends rather than absolute values, we have therefore chosen to use the
familiar ``diskbb'' model.

\begin{figure}
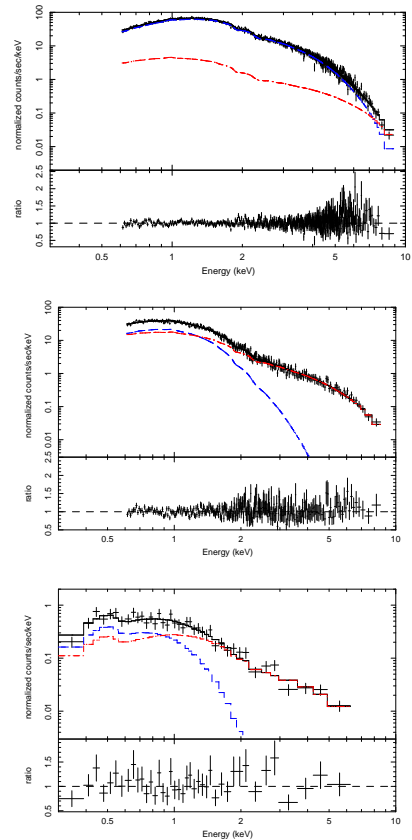

\begin{center}
\scalebox{0.5}{\rotatebox{270}{\plotone{fig2a.ps}}}\\
\scalebox{0.5}{\rotatebox{270}{\plotone{fig2b.ps}}}\\
\scalebox{0.5}{\rotatebox{270}{\plotone{fig2c.ps}}}
\caption{\label{fig:pospectra}Three XRT spectra of \xte fit with an absorbed
  optically thick, geometrically thin accretion disk, with the addition of a
  power-law component, denoted ``diskbb+po'' in the text.  The accretion disk
  component is colored blue, and the power-law component is colored red.  The
  top panel shows the spectrum from observation 01, when the accretion disk
  dominates the emission.  The middle panel shows observation 16, when the
  disk and power-law have roughly equal contributions.  The bottom panel shows
  observation 22 when the power-law component dominates.  There is clear
  evidence for both components in all three spectra.}
\end{center}
\end{figure}

In the first model, we have combined an accretion disk with a
power-law component, denoted ``diskbb+po''.  This power-law component is a
purely phenomenological fit and could be from a compact jet, a corona, or
reprocessed emission.  The results from this fit are shown in
Table~\ref{tab:diskpo}.  The best-fit $\chi^2$ per degree of freedom ($\nu$)
are acceptable for each observation.  Using this model we observe the
accretion disk cools significantly over the course of the observations from
$\sim0.8\,\mathrm{keV}$ to $\sim0.2\,\mathrm{keV}$.  Figure~\ref{fig:pospectra}
shows three sample XRT spectra from observation 01, when the accretion disk
dominates the emission; observation 16, when the disk and power-law have
roughly equal contributions; and observation 22, when the power-law component
dominates.  The accretion disk component is significant in all spectra, except
in observation 21.  This low count-rate dataset did not require a soft
component to achieve a good fit, but at the same time a disk contribution
similar to those of observations 20 and 22 is consistent with the data.
Through all the observations the normalization of the disk component, which is
proportional to the square of the inner disk radius, does not show a
significant trend.  The spectral index of the power-law component varies
considerably, from 1.4--3.2, also with no significant trend.  However, for most
of the early observations the power-law spectral index is not well constrained as
the soft disk flux dominates the emission.

\begin{deluxetable*}{cccccccc}
\tablewidth{0pt}
\tablecaption{\label{tab:diskpo}XRT Spectral fits with ``diskbb+po'' model}
\tabletypesize{\scriptsize}
\tablehead{
\colhead{Obs.} & \colhead{$kT$} & \colhead{Norm} & \colhead{$\Gamma$} &
\colhead{Norm.} & 
\colhead{Flux} &
\colhead{Disk Flux} &
\colhead{$\chi^2/\nu$}\\
 & (keV) & & &  & ($\mathrm{erg}\,\mathrm{cm}^{-2}\,\mathrm{s}^{-1}$) &
($\mathrm{erg}\,\mathrm{cm}^{-2}\,\mathrm{s}^{-1}$) &
}
\startdata
01 & 0.83(1) & $4.6(2)\times10^{3}$ & 2.0(2) & 0.8(2) & 
$4.3(2)\times10^{-8}$ & $3.9(2)\times10^{-8}$ & 632.8/513\\

02 & 0.768(4) & $3.7(1)\times10^{3}$ & 1.4(3) & $0.13^{+0.09}_{-0.07}$ & 
$2.38(9)\times10^{-8}$ & $2.26(5)\times10^{-8}$ & 1190.4/953\\

    &          & $3.4(1)\times10^{3}$ &        & $0.07^{+0.06}_{-0.04}$ &
$2.13(6)\times10^{-8}$ & $2.07(4)\times10^{-8}$ & \\

03 & 0.746(4) & $3.8(1)\times10^{3}$ & 1.6(6) & $0.10^{+0.09}_{-0.07}$ &
$2.10(8)\times10^{-8}$ & $2.03(5)\times10^{-8}$ & 906.4/881\\

    &          & $3.7(1)\times10^{3}$ &        & $0.10^{+0.09}_{-0.08}$ &
$2.02(8)\times10^{-8}$ & $1.95(5)\times10^{-8}$ & \\

04 & 0.700(7) & $3.4(2)\times10^{3}$ & 2.1(1) & 0.6(1) &
$1.61(8)\times10^{-8}$ & $1.39(7)\times10^{-8}$ & 596.1/493\\

05 & 0.694(8) & $3.5(2)\times10^{3}$ & 2.1(1) & 0.6(1) &
$1.61(9)\times10^{-8}$ & $1.39(8)\times10^{-8}$ & 571.8/448\\

06 & 0.611(8) & $3.3(2)\times10^{3}$ & 2.7(1) & 0.42(7) &
$8.6(5)\times10^{-9}$ & $7.4(5)\times10^{-9}$ & 391.7/372\\

07 & 0.646(6) & $3.2(1)\times10^{3}$ & 2.4(2) & 0.19(5) &
$9.7(4)\times10^{-9}$ & $9.1(4)\times10^{-9}$ & 848.4/751\\

    &          & $2.9(1)\times10^{3}$ &        & 0.24(7) &
$9.2(5)\times10^{-9}$ & $8.4(4)\times10^{-9}$ & \\

08 & 0.61(1)  & $2.7(2)\times10^{3}$ & 2.8(2) & 0.29(5) &
$6.8(5)\times10^{-9}$ & $6.0(4)\times10^{-9}$ & 449.8/337\\

09 & 0.611(5) & $3.5(2)\times10^{3}$ & 2.7(1) & $0.31^{+0.12}_{-0.06}$ &
$8.6(6)\times10^{-9}$ & $7.8(4)\times10^{-9}$ & 1051.7/941\\

    &          & $3.7(2)\times10^{3}$ &       & 0.33(8) &
$9.2(4)\times10^{-9}$ & $8.3(4)\times10^{-9}$ & \\

    &          & $4.2(2)\times10^{3}$ &       & $0.41^{+0.03}_{-0.06}$ &
$1.05(4)\times10^{-8}$ & $9.4(4)\times10^{-9}$ & \\

10 & 0.532(8) & $4.3(3)\times10^{3}$ & 3.0(2) & 0.32(5) &
$6.0(4)\times10^{-9}$ & $5.2(4)\times10^{-9}$ & 419.2/329\\

11 & 0.529(7) & $3.4(2)\times10^{3}$ & 2.8(2) & 0.17(4) &
$4.4(3)\times10^{-9}$ & $4.0(3)\times10^{-9}$ & 418.6/322\\

12 & 0.495(7) & $3.7(3)\times10^{3}$ & 3.2(2) & 0.20(3) &
$3.7(2)\times10^{-9}$ & $3.2(2)\times10^{-9}$ & 334.3/287\\

13 & 0.480(7) & $3.3(3)\times10^{3}$ & 3.1(1) & 0.22(5) &
$3.0(2)\times10^{-9}$ & $2.5(2)\times10^{-9}$ & 550.6/473\\

    &          & $3.5(3)\times10^{3}$ &        & 0.20(3) &
$3.1(2)\times10^{-9}$ & $2.6(2)\times10^{-9}$ & \\

15 & 0.415(7) & $3.3(4)\times10^{3}$ & 2.5(1) & 0.28(3) &
$2.2(2)\times10^{-9}$ & $1.3(1)\times10^{-9}$ & 531.8/544\\

    &          & $3.2(4)\times10^{3}$ &        & 0.30(4) &
$2.2(2)\times10^{-9}$ & $1.3(2)\times10^{-9}$ &\\

16 & 0.293(6) & $8(1)\times10^{3}$ & 2.4(1) & 0.27(3) &
$1.4(1)\times10^{-9}$ & $5.6(1)\times10^{-10}$ & 544.7/563\\

    &          & $7(1)\times10^{3}$ &        & 0.27(3) &
$1.3(1)\times10^{-9}$ & $4.8(1)\times10^{-10}$ & \\

17 & 0.21(1) & $2.3(4)\times10^{4}$ & 2.3(1) & 0.14(1) &
$7.3(7)\times10^{-10}$ & $2.4(5)\times10^{-10}$ & 436.4/415\\

    &         & $2.3(5)\times10^{4}$ &        & 0.15(2) &
$7.6(7)\times10^{-10}$ & $2.5(6)\times10^{-10}$ &\\

19 & 0.2(1) & $6^{+27}_{-5}\times10^{2}$ & 1.7(3) & $8(2)\times10^{-3}$  &
$5(1)\times10^{-11}$ & $2.7^{+12}_{-2.5}\times10^{-12}$ & 13.3/23\\

20 & 0.19(4) & $3^{+5}_{-1}\times10^{3}$ & 1.5(3) & 0.010(4) &
$9(3)\times10^{-11}$ & $1.8^{+3.2}_{-1.0}\times10^{-11}$ & 28.3/25\\

21 &         &                          & 2.3(2) & 0.012(1) &
$4.4(4)\times10^{-11}$ & & 19.3/20\\

22 & 0.20(3) & $4^{+4}_{-2}\times10^{3}$ & 2.1(3) & 0.024(9) &
$1.3(5)\times10^{-10}$ & $3.2^{+3.1}_{-1.8}\times10^{-11}$ & 35.0/37\\

\enddata 
\tablecomments{XRT spectral fits with a continuum model
consisting of an optically thick geometrically thin accretion disk combined
with a power-law component.  This model, ``diskbb+po'', is described in
Section~\ref{sec:spectra}.  Observations 1-17 have been fit from 0.6-10 keV,
and observations 19-22 have been fit from 0.3-10 keV.}
\end{deluxetable*}

In the second model, we have combined an accretion disk with a hot
optically thin Comptonizing corona~\citep{t94}, denoted ``diskbb+comptt''.  The
electron temperature of the Comptonizing corona was fixed at 50~keV, and the
seed photons were constrained to have the same temperature as the accretion
disk.  The results from this fit are shown in Table~\ref{tab:disktt}.  The
best-fit optical depth ($\tau$) of the Comptonizing corona is quite low, and
for many observations we could only obtain an upper limit on $\tau$.  Similar
to our fits with the previous model, we were unable to fit the
Comptonized model to observation 21, as we could not obtain a significant
constraint on the accretion disk to provide seed photons for the hard
component.  As above, the best-fit values of $\chi^2/\nu$ are acceptable for
each observation, and we see the accretion disk cooling with no significant
trend in the accretion disk normalization parameter and inner disk radius.

\begin{deluxetable*}{ccccccccc}
\tablewidth{0pt}
\tablecaption{\label{tab:disktt}XRT spectral fits with ``diskbb+comptt'' model}
\tabletypesize{\scriptsize}
\tablehead{
\colhead{Obs.} & \colhead{$kT$} & \colhead{Norm} & 
\colhead{$kT_e$} & \colhead{$\tau$} & \colhead{Norm.} & 
\colhead{Flux} &
\colhead{Disk Flux} &
\colhead{$\chi^2/\nu$}\\
 & (keV) & & (keV) & &  & ($\mathrm{erg}\,\mathrm{cm}^{-2}\,\mathrm{s}^{-1}$) 
& ($\mathrm{erg}\,\mathrm{cm}^{-2}\,\mathrm{s}^{-1}$) & 
}
\startdata
01 & 0.71(6) & $8(1)\times10^{3}$ & 50.0 & $<3.4\times10^{-2}$ & 0.05(2) &
$4.3(8)\times10{-8}$ & $3.4(6)\times10^{-8}$ & 626.9/513\\

02 & 0.74(3) & $4.3(4)\times10^{3}$ & & $0.12^{+0.3}_{-0.09}$ &
$9^{+6}_{-3}\times10^{-3}$ & $2.4(3)\times10^{-8}$ & $2.2(2)\times10^{-8}$ &
1173.6/953\\

    &         & $3.9(4)\times10^{3}$ & 50.0 &                       &
$5^{+5}_{-2}\times10^{-3}$ & $2.1(2)\times10^{-8}$ & $2.0(2)\times10^{-8}$ &\\

03 & 0.72(3) & $4.4(5)\times10^{3}$ & 50.0 & $<0.4$ &
$7^{+6}_{-4}\times10^{-3}$ & $2.1(3)\times10^{-8}$ & $2.0(2)\times10^{-8}$ &
899.8/881\\

    &         & $4.2(5)\times10^{3}$ &      &        &
$6^{+6}_{-4}\times10^{-3}$ & $2.0(3)\times10^{-8}$ & $1.9(2)\times10^{-8}$ &\\

04 & 0.58(5) & $7(2)\times10^{3}$ & 50.0 & $0.03(2)$ &
$3(2)\times10^{-2}$ & $1.6(4)\times10^{-8}$ & $1.2(3)\times10^{-8}$ & 576/493\\

05 & 0.56(5) & $8(2)\times10^{3}$ & 50.0 & $0.03^{+0.03}_{-0.01}$ &
$0.03(1)$ & $1.6(4)\times10^{-8}$ & $1.1(3)\times10^{-8}$ & 571.8/448\\

06 & 0.51(3) & $7(1)\times10^{3}$ & 50.0 & $<0.02$ &
$0.013(6)$ & $9(1)\times10^{-9}$ & $7(1)\times10^{-9}$ & 419.4/372\\

07 & 0.60(2) & $4.3(4)\times10^{3}$ & 50.0 & $<0.02$ &
$6(2)\times10^{-3}$ & $9.7(9)\times10^{-9}$ & $8.6(9)\times10^{-9}$ &
855.5/751\\

    &         & $4.1(4)\times10^{3}$ &  &         &
$6(2)\times10^{-3}$ & $9.1(9)\times10^{-9}$ & $8.1(7)\times10^{-9}$ &\\

08 & 0.53(3) & $4.7(4)\times10^{3}$ & 50.0 & $<0.04$ &
$7(3)\times10^{-3}$ & $6.7(7)\times10^{-9}$ & $5.7(4)\times10^{-9}$ &
476.4/337\\

09 & 0.54(1) & $5.6(5)\times10^{3}$ & 50.0 & $<0.02$ &
$8(2)\times10^{-3}$ & $8.5(7)\times10^{-9}$ & $7.3(6)\times10^{-9}$ &
1090.7/941\\

    &         & $6.0(4)\times10^{3}$ &  &          &
$9(1)\times10^{-3}$ & $9.2(6)\times10^{-9}$ & $7.9(6)\times10^{-9}$ &\\

    &         & $6.8(5)\times10^{3}$ &  &          &
$1.0(2)\times10^{-2}$ & $1.0(1)\times10^{-8}$ & $8.9(6)\times10^{-9}$ &\\

10 & 0.46(1) & $7.6(6)\times10^{3}$ & 50.0 & $<0.02$ &
$7(2)\times10^{-3}$ & $5.9(4)\times10^{-9}$ & $5.0(4)\times10^{-9}$ &
453.7/329\\

11 & 0.47(3) & $5(1)\times10^{3}$ & 50.0 & $<0.04$ &
$5^{+1}_{-3}\times10^{-3}$ & $4.4^{+0.2}_{-0.7}\times10^{-9}$ &
$3.7^{+0.2}_{-0.6}\times10^{-9}$ & 431.5/322\\

12 & 0.44(2) & $6.5(8)\times10^{3}$ & 50.0 & $<0.03$ &
$4(2)\times10^{-3}$ & $3.6^{+0.2}_{-0.5}\times10^{-9}$ &
$3.2^{+0.2}_{-0.4}\times10^{-9}$ & 369.3/287\\

13 & 0.41(2) & $7(1)\times10^{3}$ & 50.0 & $<0.02$ &
$5(2)\times10^{-3}$ & $3.0^{+0.5}_{-0.2}\times10^{-9}$ &
$2.4^{+0.4}_{-0.2}\times10^{-9}$ & 575.4/473\\

    &         & $7(1)\times10^{3}$ &  &         &
$5(2)\times10^{-3}$ & $3.1^{+0.5}_{-0.2}\times10^{-9}$ &
$2.5^{0.4}_{-0.2}\times10^{-9}$ &\\

15 & 0.33(2) & $1.0(2)\times10^{4}$ & 50.0 & 0.14(5) &
$7(2)\times10^{-3}$ & $2.2(3)\times10^{-9}$ &
$1.3(3)\times10^{-9}$ & 554.1/544\\

    &         & $1.0(2)\times10^{4}$ &  &         &
$8(2)\times10^{-3}$ & $2.2^{+0.4}_{-0.3}\times10^{-9}$ &
$1.3^{+0.3}_{-0.2}\times10^{-9}$ & \\

16 & 0.25(1) & $2.3(4)\times10^{4}$ & 50.0 & $0.27^{+0.05}_{-0.01}$ &
$8(1)\times10^{-3}$ & $1.4(2)\times10^{-9}$ &
$6.8(9)\times10^{-10}$ & 580.8/563\\

   &          & $2.1(3)\times10^{4}$ &  &                  &
$7(1)\times10^{-3}$ & $1.3(1)\times10^{-9}$ &
$6(1)\times10^{-10}$ & \\

17 & 0.19(1) & $4(1)\times10^{4}$ & 50.0 & 0.35(4) &
$5.6(6)\times10^{-3}$ & $7.3(8)\times10^{-10}$ &
$2.8(8)\times10^{-10}$ & 445/419\\

    &         & $4(1)\times10^{4}$ & &         &
$5.8(9)\times10^{-3}$ & $7.6(9)\times10^{-10}$ &
$2.8(7)\times10^{-10}$ & \\

19 & 0.16(7) & $1.4^{+66}_{-1.1}\times10^{3}$ & 50.0 & $1^{+0.8}_{-0.3}$ &
$4^{+7}_{-1}\times10^{-4}$ & $5.0^{+0.2}_{-1.2}\times10^{-11}$ &
$3.7^{+17}_{-2.7}\times10^{-12}$ & 13.8/24\\

20 & 0.18(4) & $4^{+7}_{-2}\times10^{3}$ & 50.0 & $1.3^{+1.5}_{-0.6}$ &
$6(2)\times10^{-4}$ & $9^{+5}_{-2}\times10^{-11}$ &
$2^{+4}_{-1}\times10^{-11}$ & 28.8/26 \\

22 & 0.18(5) & $8^{+16}_{-4}\times10^{3}$ & 50.0 &  $0.5(2)$ &
$1.1^{+1.1}_{-0.3}\times10^{-3}$ & $1.3^{+1.2}_{-0.3}\times10^{-10}$ &
$4^{+7}_{-2}\times10^{-11}$ & 34.3/37\\

\enddata 
\tablecomments{XRT spectral fits with a continuum model consisting of
an optically thick geometrically thin accretion disk combined with a hot
optically thin Comptonizing corona.  This model, ``diskbb+comptt'', is
described in Section~\ref{sec:spectra}.  Observations 1-17 have been fit from
0.6-10 keV, and observations 19-22 have been fit from 0.3-10 keV.}
\end{deluxetable*}

\begin{figure}
\begin{center}
\scalebox{0.9}{\rotatebox{270}{\plotone{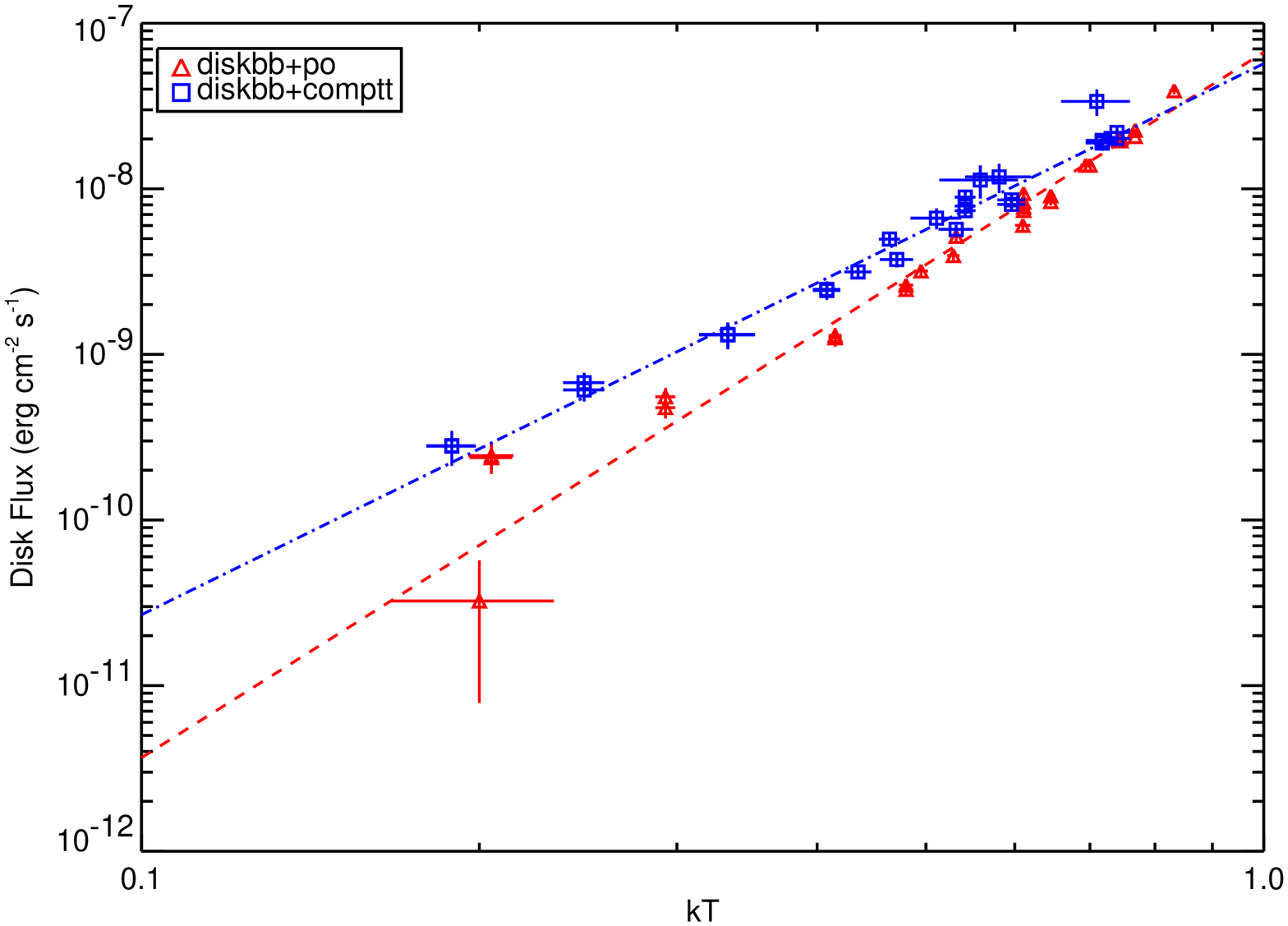}}}
\caption{\label{fig:bbody}Disk flux as a function of disk temperature for two
  different models.  The triangles represent the flux and temperature
  values from the ``diskbb+po'' model, and the dashed line represents the best
  fit to these data.  The squares represent the flux and temperature values
  from the ``diskbb+comptt'' model, and the dot-dashed line represents the best
  fit to these data.  For the ``diskbb+po'' model $f_{\mathrm{disk}}\propto
  T^{4.3 \pm 0.1}$, and for the ``diskbb+comptt'' model $f_{\mathrm{disk}}\propto
  T^{3.3 \pm 0.1}$.  In each case, for three orders of magnitude in flux the
  flux-temperature relation is very close to the predicted
  $f_{\mathrm{disk}}\propto T^{4}$.}
\end{center}
\end{figure}

Independent of our assumption for the continuum model, we see the accretion
disk cools from $\sim0.8\,\mathrm{keV}$ to $\sim0.2\,\mathrm{keV}$.  A key
theoretical prediction for a stellar-mass black hole accreting below its
Eddington limit is that $\dot{M}\propto T^{4}$ and hence
$f_{\mathrm{disk}}\propto T^{4}$~\citep*{fkr02}.  Figure~\ref{fig:bbody} shows
the disk flux as a function of disk temperature ($kT$) for each of the assumed
models.  For the ``diskbb+po'' model $f_{\mathrm{disk}}\propto
T^{4.3 \pm 0.1}$, and for the ``diskbb+comptt'' model $f_{\mathrm{disk}}\propto
T^{3.3 \pm 0.1}$.  In each case the disk flux-temperature relation is close to the
predicted relation over three orders of magnitude.  This is a further
demonstration that the size of the thin accretion disk is not changing
significantly as it cools.  The observed $T^4$ scaling also gives one some
confidence that the employed ``diskbb'' model is validated, since the data
scales as expected for such a simple model.

\subsection{X-ray and NUV Light Curve Comparison}

\begin{figure}
\begin{center}
\scalebox{1.0}{\plotone{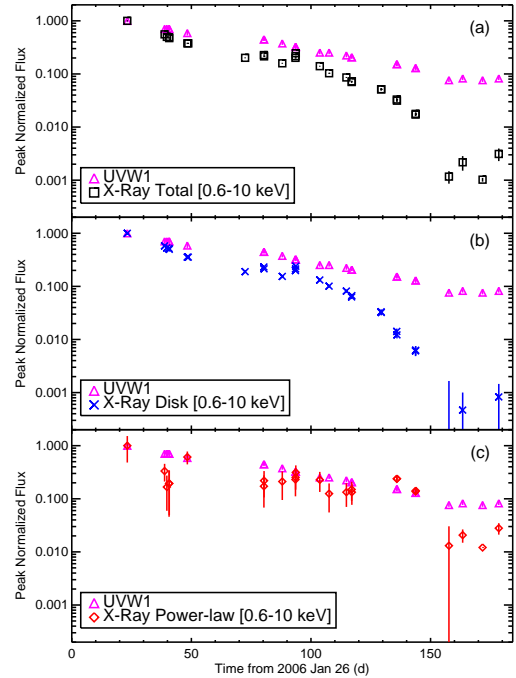}}
\caption{\label{fig:lightcurve_3panel}Comparison of the NUV $UVW1$ and
  X-ray [0.6-10 keV] light curves over the 160 days of \swift{} observations.
  The top panel (a) shows the total X-ray flux (black squares) and the $UVW1$
  flux (magenta triangles).  The middle panel (b) shows the X-ray flux due to
  the disk component (blue x's) and the $UVW1$ flux.  The bottom panel (c)
  shows the X-ray flux due to the power-law component (red diamonds) and the
  $UVW1$ flux.  In all cases the flux values have been normalized to a peak of
  1.0 at the time of the first observation; the time scale on the X axis is
  relative to the outburst date of 2006 January 26.  The NUV
  flux most closely tracks the X-ray power-law emission, and does not track the
  total X-ray flux or the X-ray disk flux.}
\end{center}
\end{figure}

We next investigated the relationship between the X-ray light curve and NUV
light curve during the decline of the outburst of \xte.  The NUV light
curve has been obtained in a single band, $UVW1$, with a peak response of 2600
\AA.  We have compared this to the X-ray flux values calculated with the
``diskbb+po'' model.  Using the flux values from the ``diskbb+comptt'' model
gives similar results.  Figure~\ref{fig:lightcurve_3panel}-a shows the
comparison of the total X-ray flux (black squares) to the $UVW1$ flux (magenta
triangles) over the 160 days of observations.  Each light curve has been
normalized to a peak of 1.0 at the time of the first observation.
Figure~\ref{fig:lightcurve_3panel}-b shows the comparison of the X-ray flux due
to the disk component (blue x's) to the $UVW1$ flux, and
Figure~\ref{fig:lightcurve_3panel}-c shows the comparison of the X-ray flux due
to the power-law component (red diamonds) to the $UVW1$ flux.  It is readily
apparent that the NUV flux most closely tracks the power-law flux in
Figure~\ref{fig:lightcurve_3panel}-c and does not track the disk flux in
Figure~\ref{fig:lightcurve_3panel}-b.  The primary source of this difference is
the $T^4$ scaling from above: the disk component of the X-ray flux declines
very rapidly, which is not apparent in the NUV light curve.  To mirror the
X-ray variation, the NUV flux would have to fade by a factor of $\sim250$
which we do not see at any time during our observations.

\subsection{\label{sec:reproc}NUV Emission as Reprocessed X-Ray Emission}

We next investigate whether the NUV light is consistent with hard X-ray
emission reprocessed by the outer accretion disk.  \citet{pm94} show that under
simple geometric assumptions reprocessed emission should be proportional to
$L_X^{0.5}\,a$, where $a$ is the orbital separation of the system.  Another
possibility, as suggested by the simultaneous spectral fitting above, is that
we see jet emission directly in the NUV.  \citet{rfhbh06} point out that if
the optical/NIR (and by extension the NUV) spectrum is jet dominated then
it should be flat from the radio regime through the optical, and
$L_{\mathrm{Opt/NIR/NUV}} \propto L_X^{0.7}$.  In a study of an ensemble of 33
BHXRBs observed over many orders of magnitude of X-ray luminosity,
\citet{rfhbh06} have shown that the optical/NIR emission is more consistent
with the prediction of reprocessed hard X-ray emission.  Our multiple
observations of \xte{} during its decline from outburst allow us to trace this
relationship for a single source over several orders of magnitude of the X-ray
luminosity and the accretion rate.

\begin{figure}
\begin{center}
\scalebox{0.9}{\rotatebox{270}{\plotone{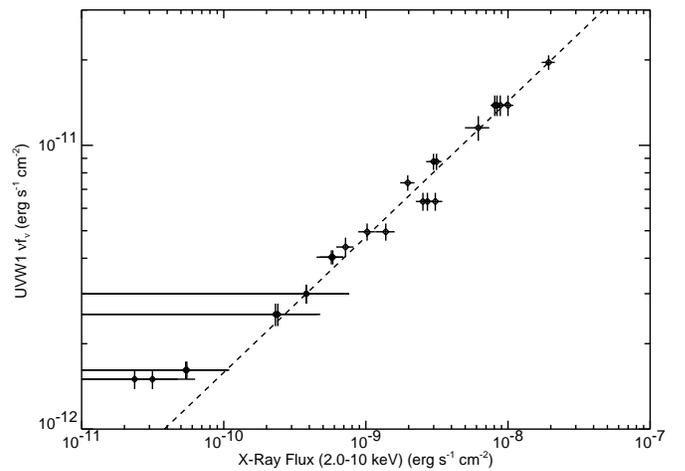}}}
\caption{\label{fig:nufnu_hard}NUV flux vs. hard X-ray flux for \xte.  The
  $UVW1$ flux tracks closely with the hard (2-10 keV) X-ray emission.  The
  best-fit slope is $0.47 \pm 0.02$, consistent with the hypothesis that the
  NUV emission is reprocessed hard X-ray emission.}
\end{center}
\end{figure}

Figure~\ref{fig:nufnu_hard} shows the NUV flux ${\nu}f_{\nu}$ for the $UVW1$
filter vs. the hard (2-10 keV) X-ray flux.  We see a very strong correlation,
with a best-fit power-law index of $0.47 \pm 0.03$, consistent with the
hypothesis that the NUV emission is dominated by reprocessed hard X-ray
emission.  It is important to note that this relationship, which is independent
of reddening in the direction of the source, holds over more than two orders of
magnitude of X-ray flux. As shown above with the light curve comparisons, this
relationship is \emph{not} consistent with the NUV emission being dominated by
direct emission from the thermal accretion disk.

\citet{kr98} predicted that when the optical/NUV flux is dominated by
reprocessed hard X-ray emission, the $e$-folding time of the optical/NUV light
curve ($\tau_{NUV}$) should be roughly twice the $e$-folding time of the X-ray
light curve ($\tau_X$), where the decay is roughly exponential: $f \sim
f_0e^{-t/\tau}$.  This relationship has been observed in several X-ray novae,
with typical X-ray light curve decay timescales of $\sim
30\,\mathrm{d}$~\citep{csl97}.  The hard X-ray decay constant of \xte{} is best
measured with the densely sampled ASM light curve shown in
Figure~\ref{fig:asmswift}.  Using the ASM data from 5 d to 150 d after the
initial outburst, when the transient was no longer significantly detected in
the 1-day averages, we find $\tau_X = 30 \pm 3\,\mathrm{d}$.  This is
consistent with the $e$-folding time measured from the XRT light curve in the
hard (2-10 keV) band, which is $\tau_X \sim 35\,\mathrm{d}$.  The NUV
light curve, using the data from 23 d to 150 d after the outburst, is well fit
with an $e$-folding time of $61 \pm 2\,\mathrm{d}$; the $e$-folding time is not
significantly different when using the entire NUV data set.  Thus the ratio of
$e$-folding times, $\tau_{NUV}/\tau_X \sim 1.7-2.0$, shows that the NUV
emission is consistent with being dominated by reprocessed hard X-ray emission
through the final \swift{} observation.

\subsection{\label{sec:simul}XRT and UVOT Simultaneous Spectral Analysis}

In order to provide additional constraints on the nature of the NUV emission,
we have performed simultaneous spectral fits with the XRT spectra and the UVOT
images.  Only the first three observations, when the disk flux dominated the
X-ray emission, were obtained with six UVOT filters. For most of the other UVOT
observations only the $UVW1$ filter was used, providing fewer constraints on
the simultaneous spectral fits.  However, even a single NUV band is sufficient
to test whether or not the NUV emission might be an extrapolation of the X-ray
disk emission or power-law emission.  We illustrate these broadband fits with
observation 01, when the accretion disk emission dominates the X-ray flux;
observation 16, when the accretion disk and hard power-law have roughly equal
contributions; and observation 22, when the power-law emission dominates.

The UVOT magnitudes were converted to XSPEC compatible files using a modified
version of ``uvot2pha'', and the latest UVOT spectral response files (v103).
We fit the XRT and UVOT spectra using the accretion disk and power-law model
(``diskbb+po'') described previously.  We fix the equivalent hydrogen column
density ($N_H$) and optical/NUV reddening with the Milky Way reddening law of
\citet{ccm89}, as described in Section~\ref{sec:nhred}.

\begin{figure}
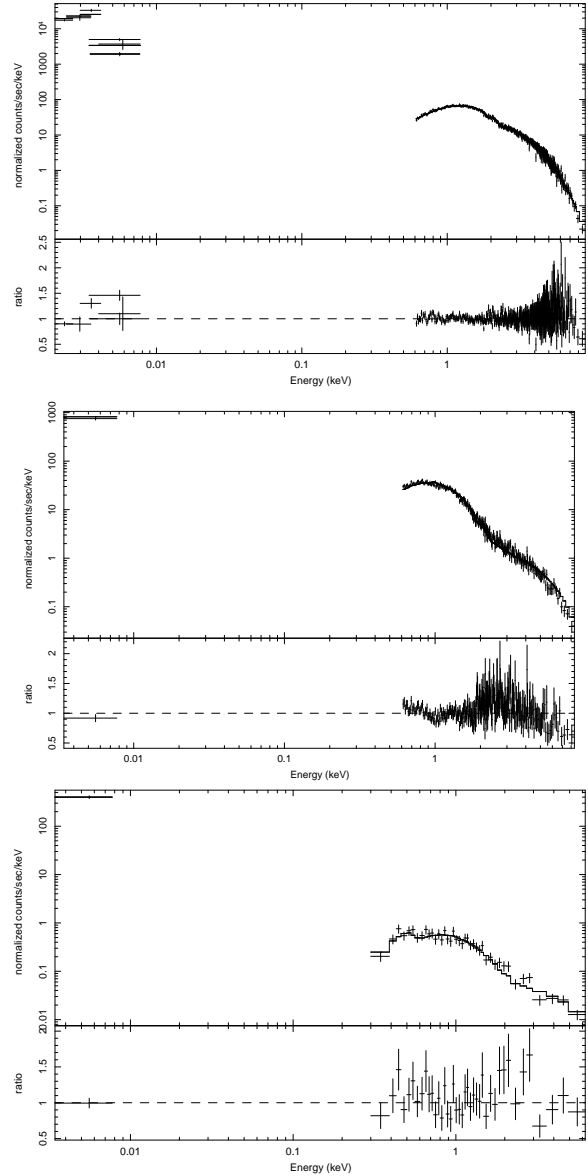

\begin{center}
\scalebox{0.7}{\rotatebox{270}{\plotone{fig6a.ps}}}\\
\scalebox{0.7}{\rotatebox{270}{\plotone{fig6b.ps}}}\\
\scalebox{0.7}{\rotatebox{270}{\plotone{fig6c.ps}}}
\caption{\label{fig:uvxrtspectra}Three simultaneous spectral fits to the XRT and
  UVOT observations of \xte.  The fit model is a reddened (in the NUV) and
  absorbed (in the X-rays) accretion disk plus an power-law component.  The
  reddening index $E(B-V)$ has been fixed at 0.215 mag.  The top panel shows
  the best fit spectrum of observation 01, when the accretion disk component
  dominates the soft X-ray emission.  The UVOT emission is well in excess of
  an extrapolation of the disk component, and is roughly consistent with an
  extrapolation of the power-law emission.  The NUV colors, which are not
  well constrained, appear a bit bluer than the power-law extrapolation
  predicts.  The middle panel shows the spectrum of observation 16.  Although
  the NUV emission is well fit by the power-law, there are large residuals
  in the X-ray spectrum.  The bottom panel shows the spectrum of observation
  22.  In this instance the NUV emission is directly on the extrapolation
  of the best-fit power-law in the X-ray spectrum.  In observation 22 an
  extrapolation of the accretion disk spectrum to the NUV provides
  negligible flux relative to the power-law.}
\end{center}
\end{figure}

The results of three spectral fits to observation 01, 16, and 22 are shown in
Table~\ref{tab:diskpouv} and plotted in Figure~\ref{fig:uvxrtspectra}.  In all
the observations, the NUV emission is well in excess of an extrapolation of the
disk flux as seen in the X-rays.  In the earliest observations the NUV emission
is a factor of $\sim3$ brighter than an extrapolation, and in the final
observations the NUV is a factor of $\sim20$ brighter.  If the Galactic
reddening is larger than we assumed, then this discrepancy is even larger.
This is consistent with our observations of the NUV and X-ray light curves,
where the X-ray disk flux fades much faster than the NUV flux.

\begin{deluxetable}{cccccc}
\tablewidth{0pt}
\tablecaption{\label{tab:diskpouv}X-ray and NUV simultaneous spectral fits}
\tabletypesize{\scriptsize}
\tablehead{
\colhead{Obs.} & \colhead{$kT$} & \colhead{Norm} & \colhead{$\Gamma$} &
\colhead{Norm.} & 
\colhead{$\chi^2/\nu$}\\
 & (keV) & & &  & 
}
\startdata
01 & 0.82(1) & $5.0(2)\times10^{3}$ & 1.49(3) & 0.32(4) & 682.2/519\\
16 & 0.32(1) & $7.9(6)\times10^{3}$ & 1.55(2) & 0.071(3) & 442.4/283\\
22 & 0.21(2) & $4(2)\times10^{3}$ & 1.73(3) & 0.016(2) & 36.5/37\\
\enddata
\tablecomments{XRT and UVOT simultaneous spectral fits with a
  continuum model consisting of an optically thick geometrically thin accretion
  disk combined with a power-law component.  This model, ``diskbb+po'', is
  described in Sections~\ref{sec:spectra} and \ref{sec:simul}.}
\end{deluxetable}

The simultaneous spectral fits in Figure~\ref{fig:uvxrtspectra} show that the
NUV excess can be adequately fit by extrapolating the hard power-law from the
X-ray band to the NUV.  This implies that the NUV light might be from the same
emission region as the hard X-ray flux.  However, there are some important
caveats.  First, the six UVOT colors in observation 01 are only marginally
consistent with the power-law extrapolation; however, the de-reddened NUV
spectral index is strongly depending on the value of $E(B-V)$ chosen, which is
not well constrained.  Second, in observation 16, where we have high
signal-to-noise and a well constrained power-law in the X-rays (as shown in the
middle panel of Figure~\ref{fig:pospectra}), $\chi^2/\nu$ of the fit is rather
poor, and significant structure is seen in the fit residuals of the middle
panel of Figure~\ref{fig:uvxrtspectra}.  It seems that using the ``diskbb+po''
model for the simultaneous spectra is overly simplistic.  Thus, the NUV
emission, while clearly in excess of the disk emission, is more consistent with
reprocessed hard X-ray emission than an extrapolation of the X-ray power-law to
the NUV regime.

\subsection{Comparison to GRO~J1655-40}

Recently, \citet{bmkgr06} obtained multiple epoch \swift{} observations of the
2005 outburst of GRO~J1655$-$40.  They observed this black hole X-ray transient
to rise from the low/hard state to the high/soft state, and for a brief time in
a very high state.  The light curve morphology of the 2005 outburst of
GRO~J1655$-$40 was much more complicated than the simple FRED profile of the
2006 outburst of \xte.  However, the spectrum of the initial observation of
GRO~J1655$-$40 closely resembles the spectrum of the final observation of \xte.
\citet{bmkgr06} find that the (0.7-9.6 keV) X-ray spectrum in this observation
is consistent with an absorbed power-law, with the addition of a low
significance relativistically broadened iron line at $\sim6.4\,\mathrm{keV}$.

We have re-analyzed the XRT observation (0003000902) of GRO~J1655$-$40 in the
low/hard state, to compare it directly to our results for \xte.  If we fit an
absorbed power-law model without a fixed absorption column density, we are able
to replicate the spectral fits from \citet{bmkgr06}.  However, the best-fit
value of $N_H = (5.9 \pm 0.2)\times10^{21}\,\mathrm{cm}^{-2}$ is significantly
lower than that obtained in the high-soft state ($N_H = 6.6 -
7.8\times10^{21}\,\mathrm{cm}^{-2}$).  By allowing the column density to float,
we are masking any contribution from a dim, soft accretion disk.  Therefore, we
have chosen to fix the column density at $6.66\times10^{21}\,\mathrm{cm}^{-2}$,
which was determined by multi-wavelength observations of the 1996 outburst of
GRO~J1655$-$40~\citep{hhsch98}.  This column density is consistent with the
values determined during the subsequent \swift{} observations, and is also
consistent with the value obtained from the Galactic HI map,
$6.9\times10^{21}\,\mathrm{cm}^{-2}$~\citep{dl90}.

\begin{figure}
\begin{center}
\scalebox{0.7}{\rotatebox{270}{\plotone{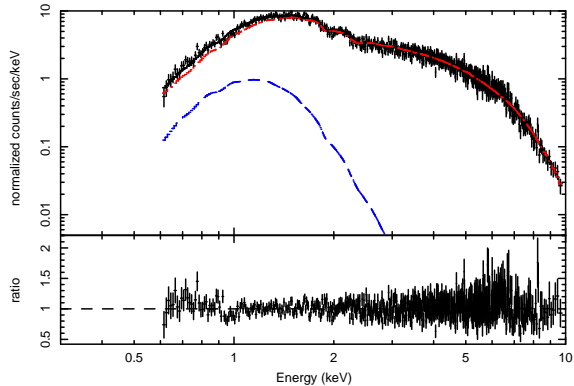}}}
\caption{\label{fig:gro1655spectrum}XRT spectrum of GRO~J1655$-$40 for observation
  02, in the low/hard state.  As in Figure~\ref{fig:pospectra}, the continuum
  model is an absorbed accretion disk with a power-law component.  The power-law
  component (red) leaves a soft excess that is consistent with a cool accretion
  disk (blue).}
\end{center}
\end{figure}

We find there is a small but significant excess of soft emission.  We first fit
the spectrum with an absorbed power-law spectrum, with a fixed column density,
and find a best fit $\chi^2/\nu = 793.4/642$.  After adding an accretion disk
component, we significantly improve the fit with a new $\chi^2/\nu =
730.4/640$.  The F-test statistic for adding the extra model component is 27.6,
with probability of $3.2\times10^{-12}$.  Therefore, the two-component
model (``diskbb+po'') is significantly better than the single
component model.  This soft excess is consistent with a thermal
accretion disk with an inner disk temperature $0.25 \pm 0.05\,\mathrm{keV}$,
and a normalization $3_{-2}^{+5}$, and unabsorbed disk flux
$8_{-3}^{+16}\times10^{-11}\,\mathrm{erg}\,\mathrm{cm}^{-2}\,\mathrm{s}^{-1}$
(0.6-10 keV).  The spectrum is shown in Figure~\ref{fig:gro1655spectrum}.  When
compared to the disk flux of GRO~J1655$-$40 in the high soft state, we find
$F_{disk}\sim T^{4}$.  \citet{bmkgr06}, in fitting for $N_H$, were unable to
see the soft spectral signature of the cool accretion disk.  However, under
reasonable assumptions about the absorption column density, we see significant
evidence for the accretion disk with an inner radius comparable to that found
in the later observations of GRO~J1655$-$40 in the high-soft state.  The X-ray
spectra of this outburst are therefore quite similar to those seen in \xte,
even with a very different light curve morphology.

\section{Discussion}

Spectral analysis of the X-ray emission from \xte{} during its 2006 outburst
shows that the geometrically thin, optically thick accretion disk which
dominates the emission in the high-soft state, is also present at or near the
innermost stable circular orbit (ISCO) in the low/hard state at very low
accretion rates.  This is consistent with recent observations of GX~339$-$4,
Cygnus X-1, and SWIFT~J1753.5$-$0127~\citep{mhsrh06,mhm06}.  In the case of
\xte, the flexible scheduling of \swift{} has made it possible, for the first
time, to densely monitor the accretion disk in both X-rays and NUV as it
cools and the source transitions from the high/soft state to the low/hard
state.  Indeed, independently of the exact form of the spectral model chosen,
the accretion disk luminosity is shown to be approximately proportional to
$T^4$, as is predicted for a blackbody of fixed size, and as seen at high
accretion rates.

At low accretion rates, the cool accretion disk does not have any significant
emission that is visible to \emph{RXTE}/PCA, which has an effective low energy
cutoff of 3~keV.  After 2006 June 9, the PCA spectral and timing signatures of
\xte{} are consistent with that of the low/hard state~\citep{rm06}, when
\swift{} observations still show a prominent accretion disk at
$\sim0.4\,\mathrm{keV}$, as is shown in the middle panel of
Figure~\ref{fig:pospectra}.  Although we do not know the distance to \xte{} or
the mass of the compact object, we can nevertheless estimate the
inner disk radius and accretion rate during the course of the observations,
scaled to a nominal distance of $d = 10\,\mathrm{kpc}$ and a black hole mass of
$M = 10\,M_\sun$.  Then the inner disk radius is $R_{in} = (6 \pm
2)\times10^{1} (d/10\,\mathrm{kpc}) / \mathrm{cos}^{-1/2}(\theta)$, where
$\theta$ is the disk inclination.  This corresponds to $R_{in} = 4 \pm 1
(M/M_\sun) (d/10\,\mathrm{kpc})/\mathrm{cos}^{-1/2}(\theta) r_g$, where $r_g =
GM/c^2$, the Schwarzchild radius, and therefore the inner radius is consistent
with the ISCO.  The corresponding luminosity during the final observation is
$L_X = (1.6 \pm 0.5) \times10^{36}
(d/10\,\mathrm{kpc})^2\,\mathrm{erg}\,\mathrm{s}^{-1}$, or $L_X/L_{Edd} = (1.2
\pm 0.4) \times 10^{-3} (M/10M_\sun) (d/10\,\mathrm{kpc})^2$, which is a small
fraction of the Eddington luminosity.

We find that the NUV (2600~\AA) light curve does not track the X-ray light
curve.  The X-ray flux, which is dominated by the thermal emission, falls much
faster than the NUV flux.  Furthermore, the NUV flux that is observed is well
in excess of an extrapolation of the X-ray disk to lower wavelengths.  Thus,
the NUV emission is not primarily due to viscous dissipation in the disk.
Meanwhile, the $e$-folding time of the NUV light curve is roughly twice the
$e$-folding time of the hard X-ray lightcurve.  This is consistent with
expectations if the optical/NUV emission is dominated by reprocessed hard X-ray
emission~\citep{kr98}.  In addition, there is a strong correlation between the
NUV flux and the hard X-ray emission, with a power-law slope of $0.47 \pm
0.03$, also consistent with reprocessing.  This relationship holds over more
than two orders of magnitude in X-ray flux, and is independent of the reddening
in the direction of the source.

The above results are consistent with the multi-wavelength observations of the
black hole X-ray transient XTE~J1859+226 during the 1999-2000 outburst.  The
optical/UV/X-ray spectral energy distribution (SED) during the early decline was
consistent with an irradiation dominated disk extending down to the
ISCO~\citep{hhcsc02}.  Studies of the optical/UV SED of several black hole
X-ray transients show that in some cases the UV emission increases towards the
far-UV (UV-hard spectra), whereas in others the emission decreases (UV-soft
spectra)~\citep{h05}.  In XTE~J1859+226 the spectrum evolved from UV-soft to
UV-hard, over five UV observations, two of which were simultaneous with X-ray
observations.  A UV-soft spectrum can be produced in a disk irradiated by a
central point source.  The UV-hard spectrum can be explained if the disk
emission is dominated by viscous dissipation or if the emission is due to X-ray
heating from an extended central source such as a jet or a
corona~(see \citeauthor{h05} \citeyear{h05} for details).  Our
results, which do not depend on the reddening in the direction of \xte, and are
based on temporal rather than spectral data, do not suffer
from this degeneracy.  The multiple epochs of simultaneous NUV and
X-ray observations made possible by \swift{} show in unprecedented detail that
the NUV emission is consistent with reprocessed hard X-ray emission in both the
high/soft and low/hard states.  This would indicate that the irradiating
central source of \xte{} is extended, which may well be typical for these black
hole X-ray transients.

In this work we have clearly shown that the NUV emission from BHXRBs may be
dominated by reprocessed hard X-ray emission.  Whereas \citet{rfhbh06} came to
this conclusion using data from many sources, we see it in detail in an
intensive study of a single source.  These observations further demonstrate
that multi-wavelength emission cannot be assumed to be direct in all cases; the
optical/NUV emission and X-ray emission may be from separate components and
separate emission regions of the same source.

These results are the strongest evidence yet obtained that accretion disks do
not automatically recede after a state transition.  Rather, the evolution of
the disk temperature appears to be smooth across state transitions, and the
inner disk appears to remain at or near the innermost stable circular orbit, at
least down to $L_X/L_{Edd}\sim0.001$.  We have made a major step forward in
being able to demonstrate this result through robust trends, while prior work
merely detected disks in the low/hard state.  Cool disks with inner radii
consistent with the ISCO have been found in every case where good quality soft
X-ray spectra have been obtained with a CCD spectrometer, including
GX~339$-$4, Cygnus X-1~\citep{mhsrh06}, SWIFT J1753.5$-$0127~\citep{mhm06},
\xte{}, and GRO~J1655$-$40 (this work).  However, we must note that geometrically thin
accretion disks are likely impossible at the lowest accretion rates observed in
black holes, and that an advective flow must take over at some point below
$L_X/L_{Edd}\sim0.001$.

\acknowledgements 

\emph{RXTE}/ASM data products were provided from the \emph{RXTE}/ASM team at
MIT.  This work has been supported in part by NSF grant AST-0407061.  DS acknowledges a
Smithsonian Astrophysical Observatory Clay Fellowship as well as support
through NASA GO grant NNG06GC05G and through the Swift Guest Investigator
program.  MAPT was supported in part by NASA LTSA grant NAG5-10889.
We thank the \swift{} operations team for their assistance with
scheduling these observations, in particular Jamie Kennea and Sally
Hunsberger, and a special thanks to Jeroen Homan for helpful discussions.

\end{document}